\documentclass[12pt]{article}
\usepackage{amsfonts}

\def\beq{\begin{equation}}
\def\eeq{\end{equation}}
\def\bea{\begin{eqnarray}}
\def\eea{\end{eqnarray}}

\def\1{\'{\i}}                           

\def\1{\'{\i}}                           
\def\k{{\kappa}}
\def\diag{{\rm\,diag\,}}

\def\Sk{{\rm\ \!S}}            
\def\Ck{{\rm\ \!C}}           
\def\Tk{{\rm\ \!T}}

\def\dd{{\rm d}}
\def\>#1{{\mathbf#1}}                 
\def\te{\theta}
\def\a{a}

 \def\pp{\pi}
\def\be{\beta}

\def\inte{Q}

\def\tea{r}

\def\tildep{{\tilde P}}
\def\tildej{{\tilde J}}

\begin{document}

\thispagestyle{empty}

\hfill \

\ 
\vspace{0.5cm}

\begin{center} {\Large{\bf{Maximal superintegrability on }}} 

{\Large{\bf{N-dimensional
curved spaces}}}

\end{center}

\bigskip\bigskip

\begin{center} Angel Ballesteros$^\dagger$, Francisco J.
Herranz$^\dagger$,\\ Mariano Santander$^\ddagger$  and
Teresa Sanz-Gil$^\ddagger$
\end{center}

\begin{center} {\it {$^\dagger$ Departamento de F\1sica,
Facultad de Ciencias,\\ Universidad de Burgos,  09001 Burgos, Spain}}
\end{center}

\begin{center} {\it {$^\ddagger$ Departamento de F\1sica Te\'orica,
Facultad de Ciencias,\\
 Universidad de Valladolid, 47011 Valladolid,
Spain}}
\end{center}

\bigskip\bigskip

\begin{abstract} 
\noindent
A  unified  algebraic  construction of
the  classical  Smorodinsky--Winternitz systems   on the
$N$D sphere, Euclidean and hyperbolic spaces through
 the Lie groups $SO(N+1)$, $ISO(N)$, and
$SO(N,1)$ is presented.  Firstly, general expressions for the
Hamiltonian and its integrals of motion are given  in a linear ambient
space $\mathbb R^{N+1}$, and secondly they are expressed in  terms of 
two       geodesic coordinate systems on the 
$N$D spaces themselves, with an explicit dependence on the 
curvature  as a parameter. On the sphere, the potential is
interpreted as a superposition of $N+1$   oscillators. 
Furthermore  each Lie algebra generator provides an integral of motion
and a set of $2N-1$ functionally independent ones are explicitly
given. In this way  the maximal superintegrability   of the $N$D
Euclidean Smorodinsky--Winternitz system is shown for any value of the
curvature. 
\end{abstract}

\newpage

\setcounter{equation}{0}

\renewcommand{\theequation}{\arabic{equation}}


\noindent
Superintegrable systems on the two- and three-dimensional
(3D) Euclidean spaces have been classified in~\cite{evans1,groscheE3}, 
and also extended to   the 2D and 3D spheres~\cite{groscheS2S3} as well
as to the hyperbolic spaces~\cite{KalninsH2,groscheH3}. Recent  
classifications of superintegrable systems for these 2D Riemannian
spaces can be found in~\cite{RS,PogosClass1,PogosClass2}. In the 2D
sphere there are two (maximal) superintegrable  potentials: the harmonic
oscillator ($\tan^2 r$)  with `centrifugal terms' and the Kepler or
Coulomb potential ($1/\tan r$)   with some `additional' terms.
 The former is the version with non-zero
curvature  of the Smorodinsky--Winternitz (SW)
system~\cite{fris,evans2,evans3,grosche1}.  Both potentials  $\tan^2 r$
and $1/\tan r$ on the $N$D sphere  
 have been studied  in  quantum mechanics   in~\cite{sch,Higgs,Leemon}, 
and have been mutually related  in~\cite{PogosDuality}.

The  SW Hamiltonian on the $N$D Euclidean space   is given by 
\beq 
 {\cal H}=\frac12\sum_{i=1}^N\left( p_i^2+ 2\be_0
q_i^2+ \frac{2\be_i}{q_i^2}\right)
\label{sw}
\eeq
  The following functions are
integrals of motion for (\ref{sw})  $(i<j;\ i,j=1,\dots,N)$:
\bea 
&&  I_{0i}=\tildep_i^2+2\be_0
q_i^2+2\frac{\be_i}{q_i^2}\qquad \mbox{with}\quad   \tildep_i=p_i
\label{za} \\ 
&& I_{ij}=\tildej_{ij}^2+2\be_i
\frac{q_j^2}{q_i^2}+ 2\be_j
\frac{q_i^2}{q_j^2}\qquad \mbox{with}\quad 
\tildej_{ij}=q_ip_j-q_jp_i  .
 \label{zb}
\eea
  The   set (\ref{za}) comes  from the separability of the Hamiltonian
$2{\cal H}=\sum_i I_{0i}$, while (\ref{zb}) are just the square of the
components of the angular momentum tensor plus some additional terms. 
The functions $\tildep_i$, $\tildej_{ij}$ close the commutation
relations of the Euclidean algebra $iso(N)$  with respect to the  
canonical Lie--Poisson bracket:
\beq
\{f,g\}=\sum_{i=1}^N\left(\frac{\partial f}{\partial q_i}
\frac{\partial g}{\partial p_i}
-\frac{\partial g}{\partial q_i} 
\frac{\partial f}{\partial p_i}\right)  .
\label{bh} 
\eeq

Our aim is to construct, simultaneously,    the
non-zero curvature version of (\ref{sw}) on   the three classical
Riemannian spaces with constant curvature  in
arbitrary dimension, as well as to prove its
maximal superintegrability, from a group theoretical standpoint.

Let ${so}_\k(N+1)$ be the real Lie algebra of the Lie group
$SO_\k(N+1)$ with generators
$\{J_{0i}\equiv P_i,J_{ij}\}$ ($i,j=1,\dots,N$; $i<j$) and  
non-vanishing commutation relations given by 
\beq 
\begin{array}{lll}
 [J_{ij},J_{ik}]=J_{jk} &\qquad  [J_{ij},J_{jk}]=-J_{ik} &\qquad 
 [J_{ik},J_{jk}]=J_{ij}  \\[2pt]
 [J_{ij},P_{i}]=P_j &\qquad  [J_{ij},P_{j}]=-P_i  &\qquad  [P_i,P_j]=\k
J_{ij} 
\end{array}
\label{aa}
\eeq
with $i<j<k$. If we consider  the following Cartan decomposition of
${so}_\k(N+1)$:
\beq
 {so}_\k(N+1)= h\oplus  p \qquad  h=\langle
J_{ij}\rangle \qquad
  p=\langle P_{i}\rangle  
\label{ac}
\eeq
where   $ h$ is the  Lie algebra
of   $H\simeq SO(N)$,  we obtain 
a family of $N$D symmetric homogeneous spaces
$ S^N_{[\k]}=SO_\k(N+1)/SO(N)$ parametrized by $\k$, which turns out
to be the constant sectional curvature of the space. Thus
  $J_{ij}$  leave a  point $\cal O$ invariant by 
acting as rotations, while    $P_i$ generate
translations that move $\cal O$ along $N$ basic  geodesics
$l_i$ orthogonal at $\cal O$.   For $\k>,=,<0$,   $ S^N_{[\k]}$
reproduces the    sphere $\>S^N=SO(N+1)/SO(N)$, Euclidean $\>E^N=
 ISO(N)/SO(N)$  and hyperbolic
 $\>H^N=SO(N,1)/SO(N)$  spaces, respectively.  The case $\k=0$ is
the contraction  around $\cal O$:
$\>S^N\rightarrow \>E^N \leftarrow\>H^N$. 
 
The  vector representation  of ${so}_\k(N+1)$  is given
by  $(N+1)\times (N+1)$ real matrices:
\beq
P_i=-\k \, e_{0i}+e_{i0} \qquad J_{ij}=- e_{ij}+e_{ji} 
\label{add}
\eeq
where $e_{ij}$ is the matrix with   entries
$(e_{ij})_m^l=\delta_i^l\delta_j^m$. Any   generator $X$ of
${so}_\k(N+1)$ fulfils 
\beq
 X^T \Lambda+\Lambda X=0 \qquad 
\Lambda= e_{00}+\k\sum_{i=1}^{N}
e_{ii}=\diag(1,\k,  \dots,  \k ) 
\label{af}
\eeq
 so that    any element $G\in 
SO_\k(N+1)$ verifies $G^T \Lambda  G=\Lambda$. 
In this way,  $SO_\k(N+1)$ is a group of linear transformations
in an ambient space $\mathbb R^{N+1}$, with   Weierstrass
coordinates  $\>x=(x_0,x_1,\dots,x_N)$, acting as the group of
isometries of the bilinear form 
$\Lambda$ via matrix multiplication. The Lie  group
$H\simeq SO(N)=\langle J_{ij}\rangle$   is the isotopy subgroup of the
origin ${\cal O}=(1,0,\dots,0)\in \mathbb R^{N+1}$. The  space $
S^N_{[\k]}$ is identified with  the orbit  of   $\cal O$, which is
contained in the `sphere' $\Sigma$:
\beq
\Sigma \ \equiv\     x_0^2+\k\sum_{i=1}^N x_i^2=1  
\label{ag}
\eeq
and the metric on $ S^N_{[\k]}$ comes 
from the flat ambient metric in $\mathbb R^{N+1}$  in the form:
\beq
\dd
s^2=\left.{1\over\k}
\left(\dd x_0^2+\k  \sum_{i=1}^{N}\dd x_i^2\right)\right|_{\Sigma} .
\label{ah}
\eeq

A  point $Q\in
S^N_{[\k]}$ with  Weierstrass coordinates $\>x $  can be reached in
different ways starting from   $\cal O$  through the action of $N$
one-parametric subgroups of $SO_\k(N+1)$:
\beq
\begin{array}{ll}
  \>x \!\!
&= \exp(\a_1 P_1)\,\exp(\a_2 P_2)\dots \exp(\a_{N-1} P_{N-1})\exp(\a_N
P_N)\, 
\cal O  \\[2pt] 
& =\exp(\te_N J_{N-1\, N})\exp(\te_{N-1} J_{N-2\, N-1})\dots\exp(\te_2
J_{12})\,
\exp(\tea P_{1})\, \cal O . 
\end{array}
\label{ba}
\eeq
The canonical parameters involved are intrinsic quantities on  
$S^N_{[\k]}$, called  
  geodesic paral\-lel $\a=(\a_1,\dots,\a_N)$ and geodesic polar
$\te=(\tea,\te_2,\dots,\te_N)$ coordinates of the point  $\>x $: 
\bea
&& {x_0=\prod_{s=1}^N\Ck_\k(\a_s)
= \Ck_\k(\tea)}\cr
&&  {x_1=\Sk_\k(\a_1)\prod_{s=2}^N\Ck_\k(\a_s)=
\Sk_\k(\tea)\cos\te_2}\label{bb}\\ 
&&  { x_i=\Sk_\k(\a_i)\prod_{s=i+1}^N\!\!\Ck_\k(\a_s)
=\Sk_\k(\tea)\prod_{s=2}^i\sin\te_s\cos\te_{i+1}}\cr
 && {  x_N=\Sk_\k(\a_N) =\Sk_\k(\tea)\prod_{s=2}^N\sin\te_s .}
\nonumber
\eea
where 
the curvature-dependent functions $\Ck_{\k}(x)$ and $\Sk_{\k}(x)$ are 
 defined by \cite{Trigo,Conf}:
\beq 
\Ck_{\k}(x) 
=\left\{
\begin{array}{ll}
  \cos {\sqrt{\k}\, x} &\   \k >0 \cr 
  1  &\  \k  =0 \cr 
\cosh {\sqrt{-\k}\, x} &\  \k <0 
\end{array}\right. \qquad 
\Sk_{\k}(x) 
=\left\{
\begin{array}{ll}
    \frac{1}{\sqrt{\k}} \sin {\sqrt{\k}\, x} &\   \k >0 \cr 
  x &\  \k  =0 \cr 
\frac{1}{\sqrt{-\k}} \sinh {\sqrt{-\k}\, x} &\  \k <0 
\end{array}\right.  
\label{ae}
\eeq
The $\k$-tangent is  defined by
$\Tk_\k(x)= {\Sk_\k(x)}/{\Ck_\k(x)}$; its  contraction $\k=0$ is 
$\Tk_0(x)=x$.  

Each parallel  coordinate  $\a_i$, associated to $P_i$, has
dimensions of {\em length}:   $\a_1$ is the distance  between $\cal
O$ and a point $Q_1$, measured along the basic geodesic $l_1$; 
$\a_2$ is the distance  between $Q_1$ and another point $Q_2$, 
 measured along a geodesic  $l'_2$ through $Q_1$ and orthogonal  
to $l_1$  (and `parallel' in the sense of parallel transport to 
$l_2$) and so on, up to reaching  $Q$ \cite{Conf}.  On the other
hand, the first polar coordinate  $\tea$, associated to
$P_1$, has dimensions of {\em length}  and is the distance  between
$\cal O$ and $Q$ measured along the geodesic $l$ joining both
points. The remaining $\te_i$, associated to $J_{i-1\,i}$,   are
ordinary {\em angles}, the  polar angles of $l$ relative
to the reference flag at $\cal O$
 spanned by  $\{l_1\}, \{l_1, l_2\}, \dots$. On the sphere
$\>S^N$ with positive curvature $\k=1/R^2$, the usual  spherical
coordinates, all of which are angles, differ from ours \cite{RS2}  
only in the first coordinate, which conventionally is taken as the 
dimensionless quantity $\tea/R$ (see for instance \cite{izmest}). When
$\k=0$ we recover  directly  the Cartesian and  polar coordinates on
$\>E^N$.

Next, by introducing  (\ref{bb})  in   (\ref{ah}),   we obtain the
  metric in $S^N_{[\k]}$:
\beq
\begin{array}{ll}
\dd s^2 \!\!&={\displaystyle{
\sum_{i=1}^{N-1} \left( \prod_{s={i+1}}^N\!\!\Ck^2_{\k}(\a_s)\right)
\dd\a_i^2+ \dd\a_N^2 }}\\[2pt]
&\displaystyle{=\dd\tea^2+\Sk^2_{\k}(\tea) \left( \dd\te_2^2+
   \sum_{i=3}^{N} 
\left( \prod_{s={2}}^{i-1} \sin^2 \te_s \right) \dd\te_i^2 \right)}  
\end{array}
\label{bc}
\eeq
which  provides  the kinetic energy $\cal T$ in terms of the
velocities ($\dot q=  \dot\a,\dot\te$), that is,    the Lagrangian  
${\cal L}\equiv{\cal T}$ of a   geodesic motion on  
$S^N_{[\k]}$.  If we introduce the   canonical
momenta $p=\partial{\cal L}/\partial{\dot q}$ ($p=p,\pp$), we obtain the
free Hamiltonian ${\cal H}\equiv{\cal T}$ on  $S^N_{[\k]}$:
\bea
{\cal T}\!\!&=&\!\!\frac 12 \left( \sum_{i=1}^{N-1}
\frac{ p_i^2 }{\prod_{s=i+1}^N \!\!\Ck_\k^2(\a_s)}    +  p_N^2
\right)\cr
\!\!&=&\!\!\frac 12 \left( \pp_1^2+ \frac{\pp_2^2}{\Sk^2_{\k}(\tea)}+
\sum_{i=3}^{N}\frac{\pp_i^2}{\Sk^2_{\k}(\tea)\prod_{s={2}}^{i-1} \sin^2
\te_s} 
\right) .
\label{bd}
\eea

 An  $N$-particle realization of  ${so}_\k(N+1)$ 
in the  phase space  is obtained by starting from the following
expressions  in terms of Weierstrass coordinates:
\beq
\tildep_{i}(x(q),\dot x(q,p))= x_0 {\dot x}_i-   x_i {\dot x}_0
 \qquad \tildej_{ij}(x(q),
\dot x(q,p))= x_i {\dot x}_j-   x_j {\dot x}_i  
\label{be}
\eeq
and expressing everything either in parallel $(\a,p)$ or polar
$(\te,\pp)$ canonical coordinates and  momenta.  In geodesic parallel
coordinates  we obtain that $(i,j=1,\dots,N)$:
\bea
&& \tildep_i=\prod_{k=1}^i\! \Ck_\k(\a_k)\Ck_\k(\a_i)p_i
+\k\Sk_\k(\a_i)\sum_{s=1}^i\Sk_\k(\a_s)
\frac{\prod_{m=1}^s \Ck_\k(\a_m)}{\prod_{l=s}^i \Ck_\k(\a_l)}\,p_s \cr
&&\tildej_{ij}=\Sk_\k(\a_i)\Ck_\k(\a_j)\prod_{s=i+1}^j\!\!  \Ck_\k(\a_s)
p_j-
\frac{\Ck_\k(\a_i)\Sk_\k(\a_j)}{\prod_{k=i+1}^j\Ck_\k(\a_k)}\,p_i
\label{bf}\\
&& \qquad +\k\Sk_\k(\a_i)\Sk_\k(\a_j)\sum_{s=i+1}^j\Sk_\k(\a_s)
\frac{\prod_{m=i+1}^s  \Ck_\k(\a_m)}{\prod_{l=s}^j  \Ck_\k(\a_l)}\,p_s  
\nonumber
\eea
 while in geodesic polar coordinates the same quantities read
$(i,j=1,\dots,N-1)$:
\bea
&&\tildep_i=\frac{\prod_{k=2}^{i+1}\sin\te_k}{\tan\te_{i+1}}\,\pp_1
+\sum_{s=2}^{i+1}
\frac{\prod_{m=s}^{i+1}\sin\te_m \, \cos\te_s\pp_s}
{\Tk_\k(\tea)\tan\te_{i+1}\prod_{l=2}^s\sin\te_l}
-\frac{\pp_{i+1}}{\Tk_\k(\tea)\prod_{l=2}^{i+1}\sin\te_l}   \cr
&&\tildep_N= \prod_{k=2}^{N}\sin\te_k \,\pp_1
+\sum_{s=2}^{N}
\frac{ \prod_{m=s}^{N}\sin\te_m \cos\te_s}
{\Tk_\k(\tea) \prod_{l=2}^s\sin\te_l}\, \pp_s  \cr
&&\tildej_{ij}= \sin\te_{i+1}\cos\te_{j+1}\!\prod_{k=i+1}^{j}\!\sin\te_k
\,\pp_{i+1} -\frac{\cos\te_{i+1}\sin\te_{j+1}}
{\prod_{l=i+1}^j\sin\te_l}\,\pp_{j+1} \label{bg}\\
&&\qquad\qquad +\cos\te_{i+1}\cos\te_{j+1}\sum_{s=i+1}^{j}
\frac{ \prod_{m=s}^{j}\sin\te_m \cos\te_s}
{  \prod_{l=i+1}^s\sin\te_l}\, \pp_s \cr
&&\tildej_{iN}= \sin\te_{i+1}\!\prod_{k=i+1}^{N}\!\sin\te_k \,\pp_{i+1}
+\cos\te_{i+1}\sum_{s=i+1}^{N}
\frac{ \prod_{m=s}^{N}\sin\te_m \cos\te_s}
{  \prod_{l=i+1}^s\sin\te_l}\, \pp_s  .
\nonumber
\eea
Both sets of  generators  (\ref{bf}) and (\ref{bg}) 
fulfil  the commutation rules (\ref{aa}) with respect to
the canonical Poisson bracket. 
 The kinetic energy is related
to the second-order Casimir of ${so}_\k(N+1)$  through
\beq
2{\cal T}= {\tilde{\cal  C}}=\sum_{i=1}^N \tildep^2_i+\k
\sum_{i,j=1}^N 
\tildej_{ij}^2
\label{bbh}
\eeq
so that any generator Poisson-commutes with ${\cal T}$. The geodesic
motion is maximally superintegrable and its   integrals of motion  
 come from any function of the Lie generators.   

Now the crucial problem is to find   potentials ${\cal U}(q)$ that
can be added to ${\cal T}$  in such a manner that the new Hamiltonian
${\cal H}={\cal T}+ {\cal U}$ preserves the maximal
superintegrability.  This   requieres to add
`some' terms to `some' functions of the generators in order
to ensure their involutivity with respect to ${\cal H}$.  By taking
into account the  results given in \cite{RS} for  
$\>S^2$ and $\>H^2$, we propose the following   generalization of
the SW potential (\ref{sw}) to the  space $S^N_{[\k]}$:  
\bea
{\cal U}\!\!&=&\!\!\be_0\frac { \sum_{s=1}^N x_s^2}{x_0^2}+ \sum_{i=1}^N
\frac {\be_i}{x_i^2}     \cr
 \!\!&=&\!\!
 \be_0 \sum_{i=1}^N\frac{\Sk_\k^2(\a_i)}{\prod_{s=1}^i
\! \Ck_\k^2(\a_s)}+\sum_{i=1}^{N-1}\frac{\be_i}
{\Sk_\k^2(\a_i)\prod_{s=i+1}^N\! \Ck_\k^2(\a_s)}+
\frac{\be_N}{\Sk_\k^2(\a_N)} \label{bj}\\
\!\!&=&\!\!\be_0
\Tk_\k^2(\tea)+\frac{1}{\Sk_\k^2(\tea)}\left(\frac{\be_1}{\cos^2\te_2}
+\sum_{i=2}^{N-1} \frac{\be_i}{ \cos^2\te_{i+1}  \prod_{s=2}^i\sin^2\te_s 
}+\frac{\be_N}{ \prod_{s=2}^N\sin^2\te_s} 
\right) .
\nonumber
\eea
  On the sphere   $\>S^N$ with $\k>0$,
this can be interpreted as the joint potential due to a
superposition of $N+1$ harmonic oscillators whose centers are
placed at $N+1$ points on  $\>S^N$ mutually separated a quadrant (a
distance $\pi/2\sqrt{\k}$, which for $\k=1$ is $\pi/2$); on
$\>S^2$ these would be placed at the three vertices of an sphere's
octant \cite{RSS}. Explicitly, if we take $\k=1$ and consider
the polar coordinate $r$ together with  $N$
geodesic distances   $r_i$ $(i=1,\dots,N)$ such that 
$x_0=\cos r$, $x_i=\cos r_i$, the potential (\ref{bj}) turns out to be
\beq
{\cal U}=\be_0\tan^2\tea +\sum_{i=1}^N
\frac{\be_i}{\cos^2 r_i}  =\be_0\tan^2\tea +\sum_{i=1}^N \be_i\tan^2 r_i
+ \sum_{i=1}^N{\be_i} . 
\eeq
The first  term is  $\be_0
\tan^2 \tea$, where  $\tea$ is the  distance from the particle and the
origin ${\cal O}$ along the geodesic $l$; this is the spherical Higgs
potential with center at ${\cal O}$ where the $0$-th
coordinate axis $x_0$ in the ambient space intersects the sphere.  Each
of the $N$ remaining terms  (apparently very different in (\ref{bj})),
$\be_i \tan^2 \tea_i$, is written  in terms of the
spherical distance $r_i$ to the point where the $i$-th coordinate axis
$x_i$ intersects the sphere. Under the contraction $\k=0$, $\>S^N\to
\>E^N$, the first term   gives rise to the `flat' harmonic oscillator
$\tea^2=\sum_i\a_i^2$, while the  $N$ remaining oscillators (whose
centers would be now `at infinity') leave the `centrifugal' barriers
$\be_i/a_i^2$ as their imprints.

Let us consider  
the following functions $I_{ij}$ 
$(i<j;\ i,j=0,1,\dots,N)$:
\beq
I_{ij}=(x_i {\dot x}_j-   x_j {\dot
x}_i)^2+2\be_i\frac{x_j^2}{x_i^2}+2\be_j\frac{x_i^2}{x_j^2}   
\label{ca}
\eeq
which are quadratic in the momenta through
  the  square of the generators.  In parallel  coordinates 
with   the phase space realization (\ref{bf}),  they turn
out to be
\bea
&& I_{0i}
=\tildep_i^2+2\be_0\,\frac{\Sk_\k^2(\a_i)}{\prod_{s=1}^i\Ck_\k^2(\a_s)}
+2\be_i\,\frac{\prod_{s=1}^i\Ck_\k^2(\a_s)}{\Sk_\k^2(\a_i)} \cr
&&
I_{ij}=\tildej_{ij}^2+2\be_i\,\frac{\Sk_\k^2(\a_j)}{
\Sk_\k^2(\a_i)\prod_{s=i+1}^j\Ck_\k^2(\a_s)}
+2\be_j\,\frac{
\Sk_\k^2(\a_i)\prod_{s=i+1}^j\Ck_\k^2(\a_s)}{\Sk_\k^2(\a_j)} .
\label{cb}
\eea
Likewise these  can be written in geodesic
polar coordinates.  Hereafter we consider the
Hamiltonian ${\cal H}={\cal T}+{\cal U}$ with ${\cal T}$ and  ${\cal
U}$ given in  (\ref{bd}) and (\ref{bj}). Notice that the
analogous property  to (\ref{bbh}) is given by
\beq
2{\cal H}= \sum_{i=1}^N I_{0i}+\k
\sum_{i,j=1}^N  I_{ij} + 2 \k\sum_{i=1}^N\be_i .
\label{ccb}
\eeq
When $\k=0$, the expressions (\ref{bf}),
(\ref{cb}) and (\ref{ccb}) reduce to (\ref{sw})--(\ref{zb}). Next it can be
proven that:
   
\noindent
{\bf Proposition 1.} {\em The $N(N+1)/2$  functions (\ref{cb}) are
integrals of the motion for   ${\cal H}$. 
}

 Let us choose the following subsets $\inte^{(k)}$  and 
$\inte_{(k)}$ of $N-1$ integrals ($k=2,\dots,N$):
\beq 
 \inte^{(k)} =\sum_{i,j=1}^k
I_{ij} \qquad  
 \inte_{(k)} =\!\!\sum_{i,j=N-k+1}^N\!\!
I_{ij}     
\eeq
where
$\inte^{(N)}\equiv\inte_{(N)}$.
  The maximal
superintegrability of $\cal H$ is characterized as follows.

\noindent
{\bf Theorem 2.} {\em (i) The   $N$  functions $\{
\inte^{(2)},\dots,\inte^{(N)},\cal H\} $ are mutually   in
involution.
The same property holds for the set    $\{
\inte_{(2)},\dots,\inte_{(N)},\cal H\} $.

\noindent
(ii) The $2N-1$ functions $\{
\inte^{(2)},\dots,\inte^{(N-1)}
,\inte^{(N)}\equiv \inte_{(N)},\inte_{(N-1)},\dots,
\inte_{(2)}, I_{0i},\cal H\}$ (with $i$ fixed) are functionally
independent, thus   
$\cal H$ is   maximally  superintegrable. }

The set  $\inte^{(k)}$  can be associated  to     a sequence of
orthogonal subalgebras  within  
$h=so(N)=\langle J_{ij}\rangle$, the   generators of which
determine the terms quadratic in the momenta in the integrals
$I_{ij}$ starting `upwards' from
$\langle J_{12}\rangle= so(2)$:
$$
\begin{array}{lllllll}
 \inte^{(2)}&\subset \inte^{(3)}&\subset  \dots &\subset  
 \inte^{(k)}&\subset 
 \dots &\subset    \inte^{(N-1)}&\subset   \inte^{(N)}\\
 {so}(2)  &\subset    {so}(3) &\subset    \dots &\subset  
 {so}(k)&  \subset  
 \dots &\subset    {so}(N-1)&  \subset {so}(N)\end{array}
$$
with a similar embedding for    
$\inte_{(k)}$  but  starting `backwards' from
$\langle J_{N-1\,N}\rangle= so(2)$. In fact, the SW
system   on $\>E^N$  can be constructed from a
coalgebra approach  \cite{BR} by means of $N$ copies of $sl(2,\mathbb
R)$. When $\k=0$, each  $\inte^{(k)}$ (or $\inte_{(k)}$) is related  to
the $k$-th order coproduct of the Casimir of $sl(2,\mathbb R)$ 
\cite{Deform}. In this sense, the results of   theorem 2 show that
the   set of integrals   ensuring the maximal superintegrability of the
`flat' SW system   coming from a 
$sl(2,\mathbb R)$-coalgebra  also holds for any curvature.

Explicit proofs and details for this algebraic  construction---which
could also be applied to  the $N$D Kepler potential---will be given
elsewhere. Furthermore, the consideration of a second contraction
parameter  $\k_2$, that determines the signature of the metric
\cite{Trigo,Conf},   would allow one to  obtain superintegrable systems
on different spacetimes.



{\section*{Acknowledgments}}

\noindent
This work was partially supported  by the Ministerio de Ciencia y
Tecnolog\1a, Spain (Projects BFM2000-1055 and  BFM2002-03773).
The authors are also grateful to G.S. Pogosyan for helpful
discussions

\bigskip



\begin{thebibliography}{40}

 \bibitem{evans1}
     Evans N W  1990
{\it   Phys. Rev. A} {\bf 41}  5666


\bibitem{groscheE3}
      Kalnins E G,  Williams G C, Miller W Jr and   Pogosyan
G S    1999 {\it J. Math. Phys. } {\bf 40}  708



\bibitem{groscheS2S3}
         Grosche C, Pogosyan G S and Sissakian A N  1995
{\it Fortschr. Phys. } {\bf 43}  523



 \bibitem{KalninsH2}
         Kalnins E G,   Miller W Jr  and  Pogosyan G S   1997
{\it J. Math. Phys. } {\bf 38}  5416



\bibitem{groscheH3}
         Grosche C, Pogosyan G S and Sissakian A N  1997
{\it Phys.  Part. Nuclei} {\bf 28}  486


 \bibitem{RS}
       Ra\~nada M F and  Santander M   1999
{\it J. Math. Phys. } {\bf 40}  5026

 \bibitem{PogosClass1}
         Kalnins E G,   Miller W Jr  and  Pogosyan G S   2000
{\it J. Phys. A: Math. Gen.} {\bf 33}  6791

 \bibitem{PogosClass2}
         Kalnins E G, Kress J M,   Pogosyan G S  and
 Miller W Jr   2001
{\it J. Phys. A: Math. Gen.} {\bf 34}  4705







 \bibitem{fris}
Fris J, Mandrosov V, Smorodinsky Ya A, Uhlir M and Winternitz P 1965
{\it   Phys. Lett.} {\bf 16}  354

 \bibitem{evans2}
     Evans N W  1990
{\it   Phys. Lett. A} {\bf 147}  483


 \bibitem{evans3}
     Evans N W  1991
{\it  J. Math. Phys.} {\bf 32}  3369

 \bibitem{grosche1}
         Grosche C, Pogosyan G S and Sissakian A N  1995
{\it Fortschr. Phys. } {\bf 43}  453












\bibitem{sch}
    Schr\"oedinger E    1940
{\it Proc. R. Ir. Acad. A} {\bf 46} 9


\bibitem{Higgs}
     Higgs P W   1979
{\it J. Phys. A: Math. Gen.} {\bf 12}  309

\bibitem{Leemon}
     Leemon H I  1979
{\it J. Phys. A: Math. Gen.} {\bf 12}  489

 

 \bibitem{PogosDuality}
         Kalnins E G,   Miller W Jr and  Pogosyan G S    2002
{\it Phys.  Atom. Nucl.} {\bf 65}  1086








\bibitem{Trigo}
      Herranz F J, Ortega R and Santander M  2000
{\it J. Phys. A: Math. Gen.} {\bf 33}  4525


\bibitem{Conf}
      Herranz F J and Santander M  2002
{\it J. Phys. A: Math. Gen.} {\bf 35}  6601


  \bibitem{RS2}
         Ra\~nada M F and  Santander M   2002
  {\it J. Math. Phys. } {\bf 43}  431


\bibitem{izmest}
         Izmest'ev A A, Pogosyan G S, Sissakian A N and Winternitz P 1999
{\it J. Math. Phys.} {\bf 40}  1549


  \bibitem{RSS} 
  Ra\~nada M F,  Santander M and Sanz-Gil T 2002 
Superintegrable potentials and the superposition of Higgs oscillators on 
  the sphere $S^2$ (Warszawa: Banach Center Publications)  to be published


\bibitem{BR}
     Ballesteros A and   Ragnisco O  1998
{\it J. Phys. A: Math. Gen.} {\bf 31}  3791



\bibitem{Deform}
     Ballesteros A and Herranz F J    1999
{\it J. Phys. A: Math. Gen.} {\bf 32}  8851



 






  
\end{thebibliography}
\end{document}